\documentclass[preprint,12pt]{elsarticle}

\usepackage{amssymb}

\journal{***}

\begin{document}

\begin{frontmatter}



\title{On a class of $(\delta+\alpha u^2)$-constacyclic codes over $\mathbb{F}_{q}[u]/\langle u^4\rangle$}


\author{Yuan Cao$^{a\ \ast}$, Yonglin Cao$^{b}$, Jian Gao$^{b}$}

\address{$^{a}$ College of Mathematics and Econometrics, Hunan University, Changsha 410082, China
 \vskip 1mm 
 $^{b}$ School of Sciences, Shandong University of Technology, Zibo, Shandong 255091, China}
\cortext[cor1]{corresponding author.  \\
E-mail addresses: yuan$_{-}$cao@hnu.edu.cn (Yuan Cao), \ ylcao@sdut.edu.cn (Yonglin Cao), \ dezhougaojian@163.com (J. Gao).}

\begin{abstract}
Let $\mathbb{F}_{q}$ be a finite field of cardinality $q$, $R=\mathbb{F}_{q}[u]/\langle u^4\rangle=\mathbb{F}_{q}+u\mathbb{F}_{q}+u^2\mathbb{F}_{q}+u^3\mathbb{F}_{q}$ $(u^4=0)$ which is a finite chain ring,
and $n$ be a positive integer satisfying ${\rm gcd}(q,n)=1$. For any $\delta,\alpha\in \mathbb{F}_{q}^{\times}$,
an explicit representation for all distinct $(\delta+\alpha u^2)$-constacyclic codes over $R$ of
length $n$ is given, and the dual code for each of these codes is determined. For the case of $q=2^m$ and $\delta=1$, all self-dual $(1+\alpha u^2)$-constacyclic codes over $R$ of
odd length $n$ are provided.
\end{abstract}


\end{frontmatter}


\section{Introduction}
\noindent
  Algebraic coding theory deals with the design of error-correcting and error-detecting codes for the reliable transmission
of information across noisy channel. The class of constacyclic codes play a very significant role in
the theory of error-correcting codes.

\par
  Let $\Gamma$ be a commutative finite ring with identity $1\neq 0$, and $\Gamma^{\times}$ be the multiplicative group of units of
$\Gamma$. For any $a\in
\Gamma$, we denote the ideal of $\Gamma$ generated by $a$ as $\langle a\rangle_\Gamma$, or $\langle a\rangle$ for simplicity, i.e., $\langle
a\rangle_\Gamma=a\Gamma=\{ab\mid b\in \Gamma\}$. For any ideal $I$ of $\Gamma$, we will identify the
element $a+I$ of the residue class ring $\Gamma/I$ with $a$ (mod $I$) for
any $a\in \Gamma$ in this paper.

\par
   A \textit{code} over $\Gamma$ of length $N$ is a nonempty subset ${\cal C}$ of $\Gamma^N=\{(a_0,a_1,\ldots$, $a_{N-1})\mid a_j\in\Gamma, \
j=0,1,\ldots,N-1\}$. The code ${\cal C}$
is said to be \textit{linear} if ${\cal C}$ is an $\Gamma$-submodule of $\Gamma^N$. All codes in this paper are assumed to be linear. The ambient space $\Gamma^N$ is equipped with the usual Euclidian inner product, i.e.,
$[a,b]_E=\sum_{j=0}^{N-1}a_jb_j$, where $a=(a_0,a_1,\ldots,a_{N-1}), b=(b_0,b_1,\ldots,b_{N-1})\in \Gamma^N$,
and the \textit{dual code} is defined by ${\cal C}^{\bot_E}=\{a\in \Gamma^N\mid [a,b]_E=0, \forall b\in {\cal C}\}$.
If ${\cal C}^{\bot_E}={\cal C}$, ${\cal C}$ is called a \textit{self-dual code} over $\Gamma$.

\par
   Let $\gamma\in \Gamma^{\times}$ and ${\cal C}$
be a linear code
over $\Gamma$ of length $N$. ${\cal C}$ is
called a $\gamma$-\textit{constacyclic code}
if $(\gamma c_{N-1},c_0,c_1,\ldots,c_{N-2})\in {\cal C}$ for all
$(c_0,c_1,\ldots,c_{N-1})\in{\cal C}$. Particularly, ${\cal C}$ is
called a \textit{negacyclic code} if $\gamma=-1$, and ${\cal C}$ is
called a  \textit{cyclic code} if $\gamma=1$.
  For any $a=(a_0,a_1,\ldots,a_{N-1})\in \Gamma^N$, let
$a(x)=\sum_{i=0}^{N-1}a_{i}x^{i}\in \Gamma[x]/\langle x^N-\gamma\rangle$. We will identify $a$ with $a(x)$ in
this paper. By [5] Propositions 2.2 and 2.4, we have

\vskip 3mm \noindent
  {\bf Lemma 1.1}  \textit{Let $\gamma\in \Gamma^{\times}$. Then ${\cal C}$ is a  $\gamma$-constacyclic code
of length $N$ over $\Gamma$ if and only if ${\cal C}$ is an ideal of
the residue class ring $\Gamma[x]/\langle x^N-\gamma\rangle$}.

\vskip 3mm \noindent
  {\bf Lemma 1.2}  \textit{The dual code of a $\gamma$-constacyclic code of length $N$ over
$\Gamma$ is a $\gamma^{-1}$-constacyclic code of length $N$ over
$\Gamma$, i.e., an ideal of $\Gamma[x]/\langle
x^N-\gamma^{-1}\rangle$}.

\vskip 3mm \par
  In this paper, let $\mathbb{F}_{q}$ be a finite field of cardinality $q$, where
$q$ is power of a prime, and denote $R=\mathbb{F}_{q}[u]/\langle u^e\rangle
=\mathbb{F}_{q}+u\mathbb{F}_{q}+\ldots+u^e\mathbb{F}_{q}$ ($u^e=0$) where $e\geq 2$. Let $e=k\geq 3$. For the case of $p=2$ and $m=1$ Abualrub and Siap [1] studied cyclic codes over the ring $\mathbb{Z}_2+u\mathbb{Z}_2$ and $\mathbb{Z}_2+u\mathbb{Z}_2+u^2\mathbb{Z}_2$ for arbitrary length $N$, then Al-Ashker and Hamoudeh [2] extended some of the results in [1],
and studied cyclic codes of an arbitrary length over
the ring $Z_2+uZ_2+u^2Z_2+\ldots+u^{k-1}Z_2$ ($u^k=0$).
For the case of $m=1$, Han  et al. [7] studied cyclic codes over $R = F_p + uF_p +\ldots+ u^{k-1}F_p$ with length $p^sn$ using discrete Fourier transform.  Singh et al. [9] studied cyclic
code over the ring $\mathbb{Z}_p[u]/\langle u^k\rangle=Z_p+uZ_p+u^2Z_p+\ldots+u^{k-1}Z_p$ for any prime integer $p$ and positive integer $N$.
Kai et al. [8] investigated $(1+\lambda u)$-constacyclic codes of arbitrary length over $\mathbb{F}_p[u]/\langle u^m\rangle$, where $\lambda$ is a unit in $\mathbb{F}_p[u]/\langle u^m\rangle$, and Cao [3] generalized
these results to $(1+w\gamma)$-constacyclic codes of arbitrary length over an arbitrary finite
chain ring $R$, where $w$ is a unit of $R$ and $\gamma$ generates the unique maximal ideal of $R$.
Sobhani et al. [10] showed that the Gray image of a $(1-u^{e-1})$-constacyclic code of length
$n$ is a length $p^{m(e-1)}n$ quasi-cyclic code of index $p^{m(e-1)-1}$.

\par
   Recently,
Sobhani [11] determined the structure of $(\delta+\alpha u^2)$-constacyclic codes
of length $p^k$ over $\mathbb{F}_{p^m}[u]/\langle u^3\rangle$, characterized and enumerated
self-dual codes among these codes, where $\delta,\alpha\in \mathbb{F}_{p^m}^{\times}$. Moreover, Sobhani proposed some open problems and further
researches in this area:

\noindent
\textsf{characterize $(\delta+\alpha u^2)$-constacyclic codes
of length $p^k$ over the finite chain ring $\mathbb{F}_{p^m}[u]/\langle u^e\rangle$ for $e\geq 4$}.

\par
   In this paper, we provide a new way different from the methods used in [7], [9] and [11] to determine the algebraic structures of
a class of $(\delta+\alpha u^2)$-constacyclic codes over the finite chain ring $\mathbb{F}_{p^m}[u]/\langle u^e\rangle$ for $e=4$.

\vskip 3mm \noindent
   {\bf Notation 1.3} Let $\delta,\alpha\in \mathbb{F}_{q}^{\times}$ and $n$ be a positive integer
satisfying ${\rm gcd}(q,n)=1$. We denote

\vskip 2mm \noindent
   $\bullet$ $R=\mathbb{F}_{q}[u]/\langle u^4\rangle=\mathbb{F}_{q}
+u\mathbb{F}_{q}+u^2\mathbb{F}_{q}+u^{3}\mathbb{F}_{q}$ ($u^4=0$).

\vskip 2mm \noindent
   $\bullet$ $\mathcal{A}=\mathbb{F}_{q}[x]/\langle(x^{n}-\delta)^2\rangle$.

\vskip 2mm \noindent
   $\bullet$ $\mathcal{A}[v]/\langle v^2-\alpha^{-1}(x^{n}-\delta)\rangle=\mathcal{A}+v\mathcal{A}$ ($v^2=\alpha^{-1}(x^{n}-\delta)$).

\vskip 3mm \par
   The present paper is organized as follows.
In Section 2, we provide an explicit representation for each $(\delta+\alpha u^2)$-contacyclic code over $R$ of length $n$ and obtain a formula to count the number of codewords in each code from its representation.
Then we give the dual code for each of such codes in Section 3. In Section 4, we determine all self-dual $(1+\alpha u^2)$-contacyclic code over $R$ of odd length $n$ for the case of $q=2^m$.
Finally, we list all $125$ distinct $(1+u^2)$-contacyclic codes over $\mathbb{F}_2[u]/\langle u^4\rangle$ of length $7$ in Section 5.




\section{Representation for $(\delta+\alpha u^2)$-constacyclic codes over $R$ of length $n$}
\noindent
In this section, we will construct a specific ring isomorphism from $\mathcal{A}+v\mathcal{A}$ onto
$R[x]/\langle x^{n}-(\delta+\alpha u^2)\rangle$. Hence we obtain a one-to-one correspondence
between the set of ideals of $\mathcal{A}+v\mathcal{A}$ onto the set of ideas of $R[x]/\langle x^{n}-(\delta+\alpha u^2)\rangle$, i.e.,
the set of $(\delta+\alpha u^2)$-constacyclic codes over $R$ of length $n$.

\par
  By Notation 1.3,
$\mathcal{A}+v\mathcal{A}=\{\xi_0+v\xi_1\mid \xi_0,\xi_1\in \mathcal{A}\}$ and the addition and multiplication are defined by

\vskip 2mm \noindent
   $(\xi_0+v\xi_1)+(\eta_0+v\eta_1)=(\xi_0+\eta_0)+v(\xi_1+\eta_1)$,

\vskip 2mm \noindent
   $(\xi_0+v\xi_1)(\eta_0+v\eta_1)=(\xi_0\eta_0+\alpha^{-1}(x^{n}-\delta)\xi_1\eta_1)+v(\xi_0\eta_1+\xi_1\eta_0)$,

\vskip 2mm \noindent
  for all $\xi_0,\xi_1,\eta_0,\eta_1\in \mathcal{A}$.

\par
  Let $\xi_0+v\xi_1\in \mathcal{A}+v\mathcal{A}$ where $\xi_0,\xi_1\in \mathcal{A}$. It is clear that
$\xi_0$ can be uniquely expressed
as $\xi_0=\xi_0(x)$ where $\xi_0(x)\in \mathbb{F}_{q}[x]$ satisfying ${\rm deg}(\xi_0(x))<2n$ (we will write ${\rm deg}(0)=-\infty$ for convenience). Dividing $\xi_0(x)$ by $\alpha^{-1}(x^{n}-\delta)$, we obtain
a unique pair $(a_0(x),a_2(x))$ of polynomials in $\mathbb{F}_{q}[x]$ such that
$$\xi_0=\xi_0(x)=a_0(x)+\alpha^{-1}(x^{n}-\delta)a_2(x), \ {\rm deg}(a_j(x))<n$$
for $j=0,2$. Similarly, there is a unique pair $(a_1(x),a_3(x))$ of polynomials in $\mathbb{F}_{q}[x]$ such that
$$\xi_1=\xi_1(x)=a_1(x)+\alpha^{-1}(x^{n}-\delta)a_3(x), \ {\rm deg}(a_j(x))<n$$
for $j=1,3$.
Denote $a_k(x)=\sum_{i=0}^{n-1}a_{i,k}x^i$ where $a_{i,k}\in \mathbb{F}_q$ for all $i=0,1,\ldots,n-1$ and
$k=0,1,2,3$. Then $\xi_0+v\xi_1$ can be uniquely written as a product of matrices:
$$\xi_0+v\xi_1=(1,x,\ldots,x^{n-1})M\left(\begin{array}{c}1 \cr v \cr \alpha^{-1}(x^{n}-\delta)\cr v\alpha^{-1}(x^{n}-\delta)\end{array}\right),$$
where $M=\left(a_{i,k}\right)_{0\leq i\leq n-1, 0\leq k\leq 3}$ is an $n\times 4$ matrix over $\mathbb{F}_{q}$. Now, we define
$$\Psi(\xi_0+v\xi_1)=(1,x,\ldots,x^{n-1})M\left(\begin{array}{c}1 \cr u \cr u^2\cr u^3\end{array}\right)=\sum_{i=0}^{n-1}\beta_ix^i,$$
where $\beta_i=\sum_{k=0}^3u^ka_{i,k}\in R$ for all $i=0,1,\ldots,n-1$. Then it is clear that $\Psi$ is a bijection from
$\mathcal{A}+v\mathcal{A}$ onto $R[x]/\langle x^n-(\delta+\alpha u^2)\rangle$. Furthermore, by
$v^2=\alpha^{-1}(x^{n}-\delta)$, $(x^{n}-\delta)^2=0$ in $\mathcal{A}+v\mathcal{A}$ and $x^n-(\delta+\alpha u^2)=0$
in $R[x]/\langle x^n-(\delta+\alpha u^2)\rangle$ one can easily verify the following conclustions.

\vskip 3mm \noindent
   {\bf Theorem 2.1} \textit{Using the notations above, $\Psi$ is a ring isomorphism from
$\mathcal{A}+v\mathcal{A}$ onto $R[x]/\langle x^n-(\delta+\alpha u^2)\rangle$}.

\vskip 3mm\noindent
  {\bf Remark} It is clear that both $\mathcal{A}+v\mathcal{A}$ and $R[x]/\langle x^n-(\delta+\alpha u^2)\rangle$ are $\mathbb{F}_q$-algebras of dimension $4n$. Specifically, $\{1,x,\ldots,x^{2n-1},v,vx,\ldots,vx^{2n-1}\}$ is an $\mathbb{F}_q$-basis of $\mathcal{A}+v\mathcal{A}$,
$\cup_{k=0}^3\{u^k,u^kx,\ldots,u^kx^{n-1}\}$  is an $\mathbb{F}_q$-basis of
$R[x]/\langle x^n-(\delta+\alpha u^2)\rangle$ and $\Psi$ is an $\mathbb{F}_q$-algebra isomorphism from
$\mathcal{A}+v\mathcal{A}$ onto $R[x]/\langle x^n-(\delta+\alpha u^2)\rangle$ determined by:
$$\Psi(x^i)=x^i \ {\rm if} \ 0\leq i\leq n-1, \ \Psi(x^n)=\delta+\alpha u^2 \ {\rm and} \ \Psi(v)=u.$$

\vskip 3mm \par
   By Theorem 2.1, in order to determine all distinct $(\delta+\alpha u^2)$-constacyclic codes over $R$ of length $n$ it is sufficient to list all distinct
ideals of $\mathcal{A}+v\mathcal{A}$.
First, we study the structures of
$\mathcal{A}$ and $\mathcal{A}+v\mathcal{A}$ in the following.

\par
   Since $\delta\in \mathbb{F}_{q}^{\times}$ and ${\rm gcd}(q,n)=1$, there are pairwise coprime monic
irreducible polynomials $f_1(x),\ldots, f_r(x)$ in $\mathbb{F}_{q}[x]$ such that
\begin{equation}
x^{n}-\delta=f_1(x)\ldots f_r(x)
\end{equation}
and $(x^{n}-\delta)^2=(x^n-\delta_0)^{2}=f_1(x)^{2}\ldots f_r(x)^{2}$.
For any integer $j$, $1\leq j\leq r$, we assume ${\rm deg}(f_j(x))=d_j$ and denote $F_j(x)=\frac{x^{n}-\delta}{f_j(x)}$.
Then $F_j(x)^2=\frac{(x^{n}-\delta)^2}{f_j(x)^{2}}$ and ${\rm gcd}(F_j(x)^2,f_j(x)^{2})=1$. Hence there exist $g_j(x),h_j(x)\in \mathbb{F}_{q}[x]$ such that
\begin{equation}
g_j(x)F_j(x)^2+h_j(x)f_j(x)^{2}=1.
\end{equation}

\par
  From now on, we adopt the following notations.

\vskip 3mm \noindent
  {\bf Notation 2.2} For any $1\leq j\leq r$, let $\varepsilon_j(x)\in \mathcal{A}$ be defined by
$$\varepsilon_j(x)\equiv g_j(x)F_j(x)^2=1-h_j(x)f_j(x)^{2} \ ({\rm mod} \ (x^{n}-\delta)^2)$$
and denote
$\mathcal{K}_j=\mathbb{F}_{q}[x]/\langle f_j(x)^{2}\rangle$.

\vskip 3mm \par
  By the Chinese remainder theorem for commutative rings, we give the structure and
properties of the ring $\mathcal{A}$.

\vskip 3mm
\noindent
  {\bf Lemma 2.3} \textit{Using the notations above, we have the following}:

\par
  (i) \textit{$\varepsilon_1(x)+\ldots+\varepsilon_r(x)=1$, $\varepsilon_j(x)^2=\varepsilon_j(x)$
and $\varepsilon_j(x)\varepsilon_l(x)=0$  in the ring $\mathcal{A}$ for all $1\leq j\neq l\leq r$}.

\par
  (ii) \textit{$\mathcal{A}=\mathcal{A}_1\oplus\ldots \oplus\mathcal{A}_r$ where $\mathcal{A}_j=\mathcal{A}\varepsilon_j(x)$ with
$\varepsilon_j(x)$ as its multiplicative identity and satisfies $\mathcal{A}_j\mathcal{A}_l=\{0\}$ for all $1\leq j\neq l\leq r$}.

\par
  (iii) \textit{For any integer $j$, $1\leq j\leq r$, and $a(x)\in \mathcal{K}_j$ we define
$\varphi_j: a(x)\mapsto \varepsilon_j(x)a(x)$ $(${\rm mod} $(x^{n}-\delta)^2)$.
Then $\varphi_j$ is a ring isomorphism from $\mathcal{K}_j$ onto $\mathcal{A}_j$}.

\par
  (iv) \textit{For any $a_j(x)\in \mathcal{K}_j$ for $j=1,\ldots,r$, define
\begin{center}
$\varphi(a_1(x),\ldots,a_r(x))=\sum_{j=1}^r\varphi_j(a_j(x))=\sum_{j=1}^r\varepsilon_j(x)a_j(x)$
\end{center}
$(${\rm mod} $(x^{n}-\delta)^2)$. Then
$\varphi$ is a ring isomorphism from $\mathcal{K}_1\times\ldots\times\mathcal{K}_r$ onto $\mathcal{A}$}.

\vskip 3mm \par
  In order to describe the structure of $\mathcal{A}+v\mathcal{A}$ ($v^2=\alpha^{-1}(x^{n}-\delta)$),
we need the following lemma.

\vskip 3mm
\noindent
  {\bf Lemma 2.4} \textit{Let $1\leq j\leq r$ and denote
$\omega_j=\alpha^{-1}F_j(x) \ ({\rm mod} \ f_j(x)^{2}).$
Then $\omega_j$ is an invertible element of $\mathcal{K}_j$ and satisfies $\alpha^{-1}(x^{n}-\delta)=\omega_jf_j(x)$ in
$\mathcal{K}_j$}.

\vskip 3mm
\noindent
  {\bf Proof.} Since $\omega_j\in \mathcal{K}_j$ satisfying $\omega_j\equiv\alpha^{-1}F_j(x)$ (mod $f_j(x)^2$),
by Equation (2) it follows that
\begin{eqnarray*}
\left(\alpha g_j(x)F_j(x)\right)\omega_j
&\equiv&\left(\alpha g_j(x)F_j(x)\right)\left(\alpha^{-1}F_j(x)\right)\\
 &=&1-h_j(x)f_j(x)^{2}\\
 &\equiv& 1 \ ({\rm mod} \ f_j(x)^{2}),
\end{eqnarray*}
which implies that $(\alpha g_j(x)F_j(x))\omega_j=1$ in the ring $\mathcal{K}_j$. Hence $\omega_j\in \mathcal{K}_j^{\times}$
and $\omega_j^{-1}=\alpha g_j(x)F_j(x)$ (mod $f_j(x)^{2}$).
By Equation (1) and $F_j(x)=\frac{x^{n}-\delta}{f_j(x)}$, we deduce that
$\alpha^{-1}(x^{n}-\delta)=\alpha^{-1}f_1(x)\ldots f_r(x)=\alpha^{-1}F_j(x)f_j(x)=
\omega_jf_j(x).$
\hfill $\Box$

\vskip 3mm
\par
   Now, we can provide the structure of $\mathcal{A}+v\mathcal{A}$.

\vskip 3mm
\noindent
  {\bf Lemma 2.5} \textit{Let $1\leq j\leq r$. Using the notations in Lemma 2.4, we denote}
$$\mathcal{K}_j[v]/\langle v^2-\omega_jf_j(x)\rangle=\mathcal{K}_j+v\mathcal{K}_j \ (v^2=\omega_jf_j(x)),$$
$$\mathcal{A}_j+v\mathcal{A}_j=\varepsilon_j(x)(\mathcal{A}+v\mathcal{A}) \ (v^2=\alpha^{-1}(x^{n}-\delta)).$$
\textit{Then we have the following conclusions}:

\vskip 2mm \par
  (i) \textit{$\mathcal{A}+v\mathcal{A}=(\mathcal{A}_1+v\mathcal{A}_1)\oplus\ldots \oplus(\mathcal{A}_r+v\mathcal{A}_r)$, where
$\varepsilon_j(x)$ is the multiplicative identity of the ring
$\mathcal{A}_j+v\mathcal{A}_j$ and this decomposition satisfies $(\mathcal{A}_j+v\mathcal{A}_j)(\mathcal{A}_l+v\mathcal{A}_l)=\{0\}$ for all $1\leq j\neq l\leq r$}.

\vskip 2mm \par
  (ii)  \textit{For any  $1\leq j\leq r$ and $a(x),b(x)\in \mathcal{K}_j$, we define
\begin{center}
$\Phi_j: a(x)+vb(x)\mapsto \varepsilon_j(x)(a(x)+vb(x))$ $(${\rm mod} $(x^{n}-\delta)^2)$.
\end{center}
Then $\Phi_j$ is a ring isomorphism from $\mathcal{K}_j+v\mathcal{K}_j$ onto $\mathcal{A}_j+v\mathcal{A}_j$}.

\vskip 2mm \par
  (iii) \textit{For any $\beta_j,\gamma_j\in \mathcal{K}_j$, $j=1,\ldots,r$, define}
$$\Phi(\beta_1+v\gamma_1,\ldots,\beta_r+v\gamma_r)=\sum_{j=1}^r\Phi_j(\beta_j+v\gamma_j)=\sum_{j=1}^r\varepsilon_j(x)(\beta_j+v\gamma_j).$$
\textit{Then
$\Phi$ is a ring isomorphism from $(\mathcal{K}_1+v\mathcal{K}_1)\times\ldots\times(\mathcal{K}_r+v\mathcal{K}_r)$
onto $\mathcal{A}+v\mathcal{A}$}.

\vskip 3mm\noindent
  {\bf Proof.} (i) Since $\varepsilon_j(x)$ is an element of
$\mathcal{A}$ and $\mathcal{A}$ is a subring of $\mathcal{A}+v\mathcal{A}$, $\varepsilon_j(x)$
is also an idempotent of the ring $\mathcal{A}+v\mathcal{A}$ for all $j=1,\ldots,r$. Then the conclusions
follow from Lemma 2.3(i) and classical ring theory.

\par
   (ii) For any $a(x),b(x)\in \mathcal{K}_j$, by the definition of $\varphi_j$ in Lemma 2.3(iii)
it follows that
\begin{eqnarray*}
\Phi_j(a(x)+vb(x))&=&(\varepsilon_j(x)a(x))+v(\varepsilon_j(x)b(x))\\
 &=&\varphi_j(a(x))+v\varphi_j(b(x)).
\end{eqnarray*}
Hence $\Phi_j$ is a bijection from $\mathcal{K}_j+v\mathcal{K}_j$ onto $\mathcal{A}_j+v\mathcal{A}_j$ by Lemma 2.3(iii).

\par
    Let $a_1(x),b_1(x),a_2(x),b_2(x)\in \mathcal{K}_j$, and denote
$\xi_i=a_i(x)+vb_i(x)\in \mathcal{K}_j+v\mathcal{K}_j$ for $i=1,2$. Since $\varphi_j$ is a ring isomorphism from $\mathcal{K}_j$ onto
$\mathcal{A}_j$, by Lemmas 2.4 and 2.5 we have that $\Phi_j(\xi_1+\xi_2)=\Phi_j(\xi_1)+\Phi_j(\xi_2)$ and
\begin{eqnarray*}
\Phi_j(\xi_1\xi_2)&=&\Phi_j((a_1(x)b_1(x)+\omega_jf_j(x)a_2(x)b_2(x))\\
   &&+v(a_1(x)b_2(x)+a_2(x)b_1(x)))\\
  &=&(\varepsilon_j(x)a_1(x))(\varepsilon_j(x)b_1(x))\\
  &&+\alpha^{-1}(x^{n}-\delta)(\varepsilon_j(x)a_2(x))(\varepsilon_j(x)b_2(x))\\
  &&+v((\varepsilon_j(x)a_1(x))(\varepsilon_j(x)b_2(x))\\
  &&+(\varepsilon_j(x)a_2(x))(\varepsilon_j(x)b_1(x)))\\
  &=&\Phi_j(\xi_1)\Phi_j(\xi_2).
\end{eqnarray*}
Therefore, $\Phi_j$ is a ring isomorphism from $\mathcal{K}_j+v\mathcal{K}_j$ onto $\mathcal{A}_j+v\mathcal{A}_j$.

\par
  (iii) It follows from (i) and (ii) immediately.
\hfill $\Box$

\vskip 3mm \par
   In order to determine all ideals of $\mathcal{A}+v\mathcal{A}$, by Lemma 2.5(iii) and
classical ring theory it is sufficient to list all distinct ideals of $\mathcal{K}_j+v\mathcal{K}_j$ ($v^2=\omega_jf_j(x)$) for
all $j=1,\ldots,r$. To do this, we need the following lemma.

\vskip 3mm
\noindent
  {\bf Lemma 2.6} (cf. [4] Example 2.1) \textit{Let $1\leq j\leq r$. Then we have the following}:

\vskip 2mm\par
  (i) \textit{$\mathcal{K}_j$ is a finite chain ring, $f_j(x)$ generates the unique
maximal ideal $\langle f_j(x)\rangle=f_j(x)\mathcal{K}_j$ of $\mathcal{K}_j$, the nilpotency index of $f_j(x)$ is equal to $2$ and the residue class field of $\mathcal{K}_j$
modulo $\langle f_j(x)\rangle$ is $\mathcal{K}_j/\langle f_j(x)\rangle\cong \mathbb{F}_{q}[x]/\langle f_j(x)\rangle$, where $\mathbb{F}_{q}[x]/\langle f_j(x)\rangle$ is an extension field of $\mathbb{F}_{q}$ with $q^{d_j}$ elements}.

\vskip 2mm\par
  (ii) \textit{Let ${\cal T}_j=\{\sum_{i=0}^{d_j-1}t_ix^i\mid t_0,t_1,\ldots,t_{d_j-1}\in \mathbb{F}_{q}\}$. Then $|{\cal T}_j|=q^{d_j}$ and every element $\xi$ of $\mathcal{K}_j$ has a unique $f_j(x)$-adic expansion:
$\xi=b_0(x)+f_j(x)b_1(x)$, $b_0(x),b_1(x)\in {\cal T}_j$.
 Hence $|\mathcal{K}_j|=|{\cal T}_j|^2=q^{2d_j}$.
Moreover, $\xi\in \mathcal{K}_j^{\times}$ if and only if $b_0(x)\neq 0$.}

\vskip 3mm \par
   Then we determine ideals of $\mathcal{K}_j+v\mathcal{K}_j$ ($v^2=\omega_jf_j(x)$).

\vskip 3mm \noindent
    {\bf Lemma 2.7} \textit{Let $1\leq j\leq r$. Then all distinct ideals of $\mathcal{K}_j+v\mathcal{K}_j$
are given by: $\langle v^l\rangle$, $l=0,1,2,3,4$. Moreover, the number of
elements contained in $\langle v^l\rangle$ is equal to $|\langle v^l\rangle|=q^{(4-l)d_j}$}.

\vskip 3mm\noindent
   {\bf Proof.} Let $\xi_0+v\xi_1\in \mathcal{K}_j+v\mathcal{K}_j$ where $\xi_0,\xi_1\in \mathcal{K}_j$.
By Lemma 2.6(ii), each $\xi_i$ has a unique $f_j(x)$-expansion: $\xi_i=b_{i,0}(x)+f_j(x)b_{i,1}(x)$,
$b_{i,0}(x),b_{i,1}(x)\in \mathcal{T}_j$, where $i=0,1$. By $f_j(x)=v^2\omega_j^{-1}$ in the
ring $\mathcal{K}_j+v\mathcal{K}_j$, it follows that

\par
$\xi_0+v\xi_1=b_{0,0}(x)+v^2\omega_j^{-1}b_{0,1}(x)$
\begin{equation}
\ \ \ \ \ \ \ \ \ \ \ \ \ \ \ \  +v\left(b_{1,0}(x)+v^2\omega_j^{-1}b_{1,1}(x)\right).
\end{equation}
By the proof of Lemma 2.4, we see that $\omega_j^{-1}=\alpha g_j(x)F_j(x)$ (mod $f_j(x)^2$). Dividing
$\omega_j^{-1}b_{i,1}(x)$ by $f_j(x)$ we obtain a unique polynomial $h_i(x)\in \mathcal{T}_j$ such that
$$\omega_j^{-1}b_{i,1}(x)=f_j(x)a_i(x)+h_i(x)$$
for some $a_i(x)\in \mathbb{F}_q[x]$. From this and by $f_j(x)^2=0$ in $\mathcal{K}_j$, we deduce that
$$v^2\omega_j^{-1}b_{i,1}(x)=v^2h_i(x)+f_j(x)\cdot f_j(x)a_i(x)=v^2h_i(x), \ i=0,1.$$
Then by (3) we obtain the following $v$-expansion for $\xi_0+v\xi_1$:
$$\xi_0+v\xi_1=b_{0,0}(x)+vb_{1,0}(x)+v^2h_0(x)+v^3h_1(x).$$
Obviously, $v^4=(v^2)^2=\omega_j^2f_j(x)^2=0$ and $v^3=v\omega_j(x)f_j(x)\neq 0$. Hence the nilpotency index of $v$ is equal to
$4$ in $\mathcal{K}_j+v\mathcal{K}_j$. Moreover, by Lemma 2.6(ii) we see that $\xi_0+v\xi_1$
is invertible if and only if $b_{0,0}(x)\neq 0$.

\par
   As stated above, we conclude that $v$ generates the unique maximal ideal
$\langle v\rangle$ of $\mathcal{K}_j+v\mathcal{K}_j$ and the residue class field
is $(\mathcal{K}_j+v\mathcal{K}_j)/\langle v\rangle=\{b_{0,0}(x)+\langle v\rangle\mid b_{0,0}(x)\in \mathcal{T}_j\}$
satisfying $|(\mathcal{K}_j+v\mathcal{K}_j)/\langle v\rangle|=|\mathcal{T}_j|=q^{d_j}$. Therefore, all distinct
ideals of $\mathcal{K}_j+v\mathcal{K}_j$ are given by:
$$\{0\}=\langle v^4\rangle\subset \langle v^3\rangle\subset\langle v^2\rangle\subset\langle v\rangle\subset
\langle v^0\rangle=\mathcal{K}_j+v\mathcal{K}_j.$$
Furthermore, for any $0\leq l\leq 4$ we have that
$$\langle v^l\rangle=\{\sum_{k=l}^4v^kt_k(x)\mid
t_k(x)\in \mathcal{T}_j, \ k=l,\ldots,3\},$$ which implies
$|\langle v^l\rangle|=|\mathcal{T}_j|^{4-l}=q^{(4-l)d_j}$.
\hfill $\Box$

\vskip 3mm \par
    Since $\Psi$ is a ring isomorphism from $\mathcal{A}+v\mathcal{A}$ onto
$R[x]/\langle x^{n}-(\delta+\alpha u^2)\rangle$, by Lemma 2.3(i) and the definition of $\Psi$ we deduce the following
corollary.

\vskip 3mm \noindent
   {\bf Corollary 2.8} \textit{For any integer $j$, $1\leq j\leq r$, denote $e_j(x)=\Psi(\varepsilon_j(x))\in R[x]/\langle x^{n}-(\delta+\alpha u^2)\rangle$. Then}

\vskip 2mm\par
   (i)  \textit{$e_1(x)+\ldots+e_r(x)=1$, $e_j(x)^2=e_j(x)$
and $e_j(x)e_l(x)=0$  in the ring $R[x]/\langle x^{n}-(\delta+\alpha u^2)\rangle$ for all $1\leq j\neq l\leq r$}.

\vskip 2mm\par
   (ii) \textit{If $\varepsilon_j(x)=e_{j,0}(x)+\alpha^{-1}(x^n-\delta)e_{j,1}(x)$ where
$e_{j,i}(x)\in \mathbb{F}_q[x]$ satisfying ${\rm deg}(e_{j,i}(x))\leq n-1$ for $i=0,1$,
then $e_j(x)=e_{j,0}(x)+u^2e_{j,1}(x)$}.

\vskip 3mm \par
    Finally, we give a precise representation for any
$(\delta+\alpha u^2)$-constacyclic code over $R$ of length $n$.

\vskip 3mm \noindent
   {\bf Theorem 2.9} \textit{Using the notations above, all distinct $(\delta+\alpha u^2)$-constacyclic codes over $R$ of length $n$
are given by}:
$$\mathcal{C}_{(l_1,\ldots,l_r)}=\left\langle \sum_{j=1}^ru^{l_j}e_j(x)\right\rangle,
\ 0\leq l_1,\ldots,l_r\leq 4.$$
\textit{Moreover, the number of codewords contained in $\mathcal{C}_{(l_1,\ldots,l_r)}$ is equal to
$|\mathcal{C}_{(l_1,\ldots,l_r)}|=q^{\sum_{j=1}^r(4-l_j)d_j}$.}

\par
  \textit{Therefore, the number of $(\delta+\alpha u^2)$-constacyclic codes over $R$ of length $n$
is equal to $5^r$}.

\vskip 3mm \noindent
   {\bf Proof.} By Theorem 2.1 and Lemma 2.5(iii), we see that $\Psi\circ \Phi$ is a ring isomorphism from
$(\mathcal{K}_1+v\mathcal{K}_1)\times\ldots\times(\mathcal{K}_r+v\mathcal{K}_r)$ onto $R[x]/\langle x^n-(\delta+\alpha u^2)\rangle$.
Let $\mathcal{C}$ be an ideal of $R[x]/\langle x^n-(\delta+\alpha u^2)\rangle$. By classical ring theory and Lemma 2.7, for any
integer $j$, $1\leq j\leq r$, there is a unique ideal $\langle v^{l_j}\rangle$ of $\mathcal{K}_j+v\mathcal{K}_j$, where $0\leq l_j\leq 4$, such that
\begin{eqnarray*}
\mathcal{C}&=&(\Psi\circ \Phi)\left(\langle v^{l_1}\rangle\times\ldots\times \langle v^{l_r}\rangle\right)\\
     &=&\Psi\left(\Phi\{(\xi_1,\ldots,\xi_r)\mid \xi_j\in \langle v^{l_j}\rangle, \ j=1,\ldots,r\}\right)\\
     &=&\Psi\left(\left\{\sum_{j=1}^r\varepsilon_j(x)\xi_j\mid \xi_j\in \langle v^{l_j}\rangle, \ j=1,\ldots,r\right\}\right)\\
     &=&\Psi\left(\bigoplus_{j=1}^r\left\langle\varepsilon_j(x) v^{l_j}\right\rangle\right)
     =\bigoplus_{j=1}^r\left\langle\Psi(\varepsilon_j(x) v^{l_j})\right\rangle.
\end{eqnarray*}
Since $\Psi\circ \Phi$ is a ring isomorphism from
$(\mathcal{K}_1+v\mathcal{K}_1)\times\ldots\times(\mathcal{K}_r+v\mathcal{K}_r)$ onto $R[x]/\langle x^n-(\delta+\alpha u^2)\rangle$,
by Lemma 2.7 we have $|\mathcal{C}|=|\langle v^{l_1}\rangle\times\ldots\times \langle v^{l_r}\rangle|
$ $=\prod_{j=1}^r|\langle v^{l_j}\rangle|=q^{\sum_{j=1}^r(4-l_j)d_j}$.

\par
   By Corollary 2.8 and the remark after Theorem 2.1, we deduce that
$\Psi(\varepsilon_j(x) v^{l_j})=\Psi(\varepsilon_j(x))\Psi(v^{l_j})=u^{l_j}e_j(x)$ for all $j=1,\ldots,r$,
which implies
$$\mathcal{C}=\bigoplus_{j=1}^r\left\langle u^{l_j}e_j(x)\right\rangle
=\langle u^{l_j}e_1(x),\ldots,u^{l_r}e_r(x) \rangle.$$
From this and by Corollary 2.8(i), one can
easily verify that $\mathcal{C}=\langle  u^{l_1}e_1(x)+\ldots+u^{l_r}e_r(x)\rangle$.

\par
   As stated above, we conclude that the number of $(\delta+\alpha u^2)$-constacyclic codes over $R$ of length $n$
is equal to $5^r$ by Lemma 2.7.
\hfill $\Box$


\section{Dual codes of $(\delta+\alpha u^2)$-constacyclic codes ${\cal C}$ over $R$
of length $n$} \label{}
\noindent
   In this section, we give the dual code ${\cal C}^{\bot_E}$ of any $(\delta+\alpha u^2)$-constacyclic code ${\cal C}$ over $R=\mathbb{F}_q[u]/\langle u^4\rangle$
of length $n$, where $\delta, \alpha\in \mathbb{F}_q^{\times}$ and ${\rm gcd}(q,n)=1$.
By Lemma 1.2, we know that ${\cal C}^{\bot_E}$ is a $(\delta+\alpha u^2)^{-1}$-constacyclic code over $R$
of length $n$, i.e., ${\cal C}^{\bot_E}$ is an ideal of the ring $R[x]/\langle x^n-(\delta+\alpha u^2)^{-1}\rangle$.

\par
   In the ring $R[x]/\langle x^n-(\delta+\alpha u^2)^{-1}\rangle$, we have $x^n=(\delta+\alpha u^2)^{-1}$, i.e., $x^{-n}=\delta+\alpha u^2$ or $(\delta+\alpha u^2)x^n=1$, which implies
\begin{equation}
x^{-1}=(\delta+\alpha u^2)x^{n-1} \ {\rm in} \ R[x]/\langle x^n-(\delta+\alpha u^2)^{-1}\rangle.
\end{equation}

\vskip 3mm \noindent
   {\bf Lemma 3.1} \textit{Define a map $\tau:R[x]/\langle x^n-(\delta+\alpha u^2)\rangle\rightarrow R[x]/\langle x^n-(\delta+\alpha u^2)^{-1}\rangle$ by the rule that
\begin{equation}
\tau(a(x))=a(x^{-1})=\sum_{i=0}^{n-1}a_ix^{-i}=(\delta+\alpha u^2)\sum_{i=0}^{n-1}a_ix^{n-i},
\end{equation}
for all $a(x)=\sum_{i=0}^{n-1}a_ix^{i}\in R[x]/\langle x^n-(\delta+\alpha u^2)\rangle$ with $a_0,a_1,\ldots,a_{n-1}\in R$.
Then $\tau$ is a ring isomorphism from $R[x]/\langle x^n-(\delta+\alpha u^2)\rangle$ onto $R[x]/\langle x^n-(\delta+\alpha u^2)^{-1}\rangle$}.

\vskip 3mm \noindent
   {\bf Proof.} For any $g(x)\in R[x]$, we define
$$\tau_0(g(x))=g((\delta+\alpha u^2)x^{n-1})=g(x^{-1}) \ ({\rm mod} \ x^n-(\delta+\alpha u^2)^{-1}).$$
Then by Equation (4), we see that $\tau_0$ is a well-defined ring homomorphism from $R[x]$ to
$R[x]/\langle x^n-(\delta+\alpha u^2)^{-1}\rangle$. For any $h(x)=\sum_{i=0}^{n-1}h_ix^i\in R[x]/\langle x^n-(\delta+\alpha u^2)^{-1}\rangle$,
we select $g(x)=(\delta+\alpha u^2)^{-1}\sum_{i=0}^{n-1}h_ix^{n-i}\in R[x]$. Then by (4) and the definition of
$\tau_0$, it follows that
\begin{eqnarray*}
\tau_0(g(x))&=&(\delta+\alpha u^2)^{-1}\sum_{i=0}^{n-1}h_i((\delta+\alpha u^2)x^{n-1})^{n-i}\\
    &=&x^n\sum_{i=0}^{n-1}h_ix^{i-n}=h(x).
\end{eqnarray*}
Hence $\tau_0$ is surjective. Then from
$$\tau_0(x^n-(\delta+\alpha u^2))=x^{-n}-(\delta+\alpha u^2)=(\delta+\alpha u^2)-(\delta+\alpha u^2)=0$$
in $R[x]/\langle x^n-(\delta+\alpha u^2)^{-1}\rangle$ and by classical ring theory, we deduce that
the map $\tau$ induced by $\tau_0$, which is defined by (5), is a surjective ring homomorphism from $R[x]/\langle x^n-(\delta+\alpha u^2)\rangle$
onto $R[x]/\langle x^n-(\delta+\alpha u^2)^{-1}\rangle$. Moreover, it is clear that
$|R[x]/\langle x^n-(\delta+\alpha u^2)\rangle|=|R|^n=|R[x]/\langle x^n-(\delta+\alpha u^2)^{-1}\rangle|$. Therefore,
$\tau$ is a bijection and hence a ring isomorphism.
\hfill $\Box$

\vskip 3mm \noindent
   {\bf Lemma 3.2} \textit{For any $a=(a_0,a_1,\ldots,a_{n-1})\in R^n$ and $b=(b_0,b_1,\ldots,b_{n-1})\in R^n$, denote
$$a(x)=\sum_{i=0}^{n-1}a_ix^i\in R[x]/\langle x^n-(\delta+\alpha u^2)\rangle,$$
$$b(x)=\sum_{i=0}^{n-1}b_ix^i\in R[x]/\langle x^n-(\delta+\alpha u^2)^{-1}\rangle.$$
Then $[a,b]_E=\sum_{i=0}^{n-1}a_ib_i=0$ if $\tau(a(x))\cdot b(x)=0$ in $R[x]/\langle x^n-(\delta+\alpha u^2)^{-1}\rangle$}.

\vskip 3mm \noindent
  {\bf Proof.} By Equation (5) and $x^n=(\delta+\alpha u^2)^{-1}$ in $R[x]/\langle x^n-(\delta+\alpha u^2)^{-1}\rangle$, it follows
that $\tau(a(x))\cdot b(x)=[a,b]_E+\sum_{i=1}^{n-1}c_ix^i$ for some $c_1,\ldots,c_{n-1}\in R$. Hence $[a,b]_E=0$ if $\tau(a(x))\cdot b(x)=0$ in $R[x]/\langle x^n-(\delta+\alpha u^2)^{-1}\rangle$.
\hfill $\Box$

\vskip 3mm \noindent
  {\bf Remark} For any $(\delta+\alpha u^2)$-constacyclic code $\mathcal{C}$ over $R$ of length $n$,  by Lemma 3.2 it follows that
$\{b(x)\in R[x]/\langle x^{n}-(\delta+\alpha u^2)^{-1}\rangle\mid \tau(a(x))\cdot b(x)=0,
\forall a(x)\in \mathcal{C}\}\subseteq\mathcal{C}^{\bot_E}.$

\vskip 3mm \par
  Now, we determine the dual code of each $(\delta+\alpha u^2)$-constacyclic code
over $R$ of length $n$.

\vskip 3mm \noindent
  {\bf Theorem 3.3} \textit{Let $\mathcal{C}=\langle \sum_{j=1}^r u^{l_j}e_j(x)\rangle$
be a $(\delta+\alpha u^2)$-constacyclic code
over $R$ of length $n$ given by Theorem 2.9. Then the dual code of
$\mathcal{C}$ is given by}:
$$\mathcal{C}^{\bot_E}=\left\langle \sum_{j=1}^r u^{4-l_j}e_j(x^{-1})\right\rangle,$$
\textit{which is an ideal of the ring $R[x]/\langle x^n-(\delta+\alpha u^2)^{-1}\rangle$}.

\vskip 3mm \noindent
  {\bf Proof.} Let $\mathcal{D}=\langle \sum_{j=1}^r u^{4-l_j}e_j(x)\rangle$ be
the ideal of $R[x]/\langle x^n-(\delta+\alpha u^2)\rangle$ generated by $\sum_{j=1}^r u^{4-l_j}e_j(x)$.
Since $\tau$ is a ring isomorphism from $R[x]/\langle x^n-(\delta+\alpha u^2)\rangle$ onto $R[x]/\langle x^n-(\delta+\alpha u^2)^{-1}\rangle$,
$\tau(\mathcal{D})$ is an ideal of $R[x]/\langle x^n-(\delta+\alpha u^2)^{-1}\rangle$. From this and
by Theorem 2.9, we deduce that
\begin{equation}
|\tau(\mathcal{D})|=|\mathcal{D}|=q^{\sum_{j=1}^r(4-(4-l_j))d_j}=q^{\sum_{j=1}^rl_jd_j}.
\end{equation}
Moreover, by
Corollary 2.8(i) it follows that
\begin{eqnarray*}
\tau(\mathcal{C})\cdot \tau(\mathcal{D})&=& \tau(\mathcal{C}\cdot \mathcal{D})\\
  &=&\tau\left(\langle \sum_{j=1}^r u^{l_j}e_j(x)\rangle\cdot\langle \sum_{j=1}^r u^{4-l_j}e_j(x)\rangle\right)\\
  &=&\tau\left(\left\langle(\sum_{j=1}^r u^{l_j}e_j(x))(\sum_{j=1}^r u^{4-l_j}e_j(x))\right\rangle\right)\\
  &=&\tau\left(\left\langle\sum_{j=1}^r (u^{l_j}u^{4-l_j})e_j(x)\right\rangle\right)\\
  &=&\tau(\{0\})=\{0\}.
\end{eqnarray*}
From this and by Lemma 3.2, we deduce that $\tau(\mathcal{D})\subseteq \mathcal{C}^{\bot_E}$. Furthermore, by (6)
and Theorem 2.9 it follows that
$$|\mathcal{C}||\tau(\mathcal{D})|=q^{\sum_{j=1}^r(4-l_j)d_j}q^{\sum_{j=1}^rl_jd_j}=q^{4\sum_{j=1}^rd_j}=q^{4n}=|R|^n.$$
Hence we conclude that $\mathcal{C}^{\bot_E}=\tau(\mathcal{D})$ since $R$ is a finite chain ring (cf. [6]).
   Finally, since $\tau$ is a ring isomorphism defined in Lemma 3.1, we see that
$\mathcal{C}^{\bot_E}=\langle \tau(\sum_{j=1}^r u^{4-l_j}e_j(x))\rangle=\langle \sum_{j=1}^r u^{4-l_j}\tau(e_j(x))\rangle
=\langle \sum_{j=1}^r u^{4-l_j}e_j(x^{-1})\rangle$.
\hfill $\Box$


\section{Self-dual $(1+\alpha u^2)$-constacyclic codes over $\mathbb{F}_{2^m}+u\mathbb{F}_{2^m}+u^2\mathbb{F}_{2^m}+u^3\mathbb{F}_{2^m}$ of odd length} \label{}
\noindent
  In this section, let $q=2^m$ where $m$ is a positive integer, $R=\mathbb{F}_{2^m}[u]/\langle u^4\rangle$, $n$ be an odd positive integer
and $\alpha\in \mathbb{F}_{2^m}^{\times}$. As $(1+\alpha u^2)^{-1}=1+\alpha u^2$ in $R$, the dual code of every
$(1+\alpha u^2)$-constacyclic code over $R$ of length $n$ is also a $(1+\alpha u^2)$-constacyclic code over $R$ of length $n$ by Lemma 1.2.

\par
   Morover, by $(1+\alpha u^2)^{-1}=1+\alpha u^2$ in $R$ and Lemma 3.1, we see that the map $\tau$ defined
by $\tau(a(x))=a(x^{-1})$ ($\forall a(x)\in R[x]/\langle x^n-(1+\alpha u^2)\rangle$) is a ring automorphism on $R[x]/\langle x^n-(1+\alpha u^2)\rangle$ satisfying $\tau^{-1}=\tau$. From this and by Corollary 2.8(i), we deduce that
for each integer $j$, $1\leq j\leq r$, there is a unique integer $j^{\prime}$, $1\leq j\leq r$, such that
\begin{center}
$\tau(e_j(x))=e_j(x^{-1})=e_{j^{\prime}}(x)$ (mod $x^n-(1+\alpha u^2)$).
\end{center}
Hence the ring automorphism $\tau$ on $R[x]/\langle x^n-(1+\alpha u^2)\rangle$ induces
a permutation $j\mapsto j^{\prime}$ on the set $\{1,\ldots,r\}$. In order to simplify notations,
we still denote this bijection
by $\tau$, i.e.,
\begin{equation}
\tau(e_j(x))=e_j(x^{-1})=e_{\tau(j)}(x) \ ({\rm mod} \ x^n-(1+\alpha u^2)).
\end{equation}
Since the permutation $\tau$ on $\{1,\ldots,r\}$ satisfies $\tau^{-1}=\tau$,
After a suitable rearrangement of $e_1(x),\ldots,e_r(x)$, there are nonnegative integers $\rho,\epsilon$ such that
$\rho+2\epsilon=r$ and

\noindent
$\bullet$  $\tau(j)=j$, $1\leq j\leq\rho$;

\noindent
$\bullet$  $\tau(\rho+i)=\rho+\epsilon+i$  and
$\tau(\rho+\epsilon+i)=\rho+i$, $1\leq i\leq\epsilon$.

\vskip 2mm\par
  Now, by Theorem 3.3
and Equation (7),
we obtain the following corollary immediately.

\vskip 3mm\noindent
   {\bf Corollary 4.1} \textit{Let $\mathcal{C}=\langle \sum_{j=1}^r u^{l_j}e_j(x)\rangle$
be a $(1+\alpha u^2)$-constacyclic code
over $R$ of length $n$ given by Theorem 2.9. Then}
$\mathcal{C}^{\bot_E}=\left\langle \sum_{j=1}^r u^{4-l_j}e_{\tau(j)}(x)\right\rangle.$

\vskip 3mm\par
   Finally, we list all self-dual $(1+\alpha u^2)$-constacyclic codes
over $R$ of length $n$ as follows.

\vskip 3mm\noindent
   {\bf Theorem 4.2} \textit{All distinct
self-dual $(1+\alpha u^2)$-constacyclic codes of length $n$ over $R$ are given by}:
$$\left\langle\sum_{j=1}^\rho u^2e_j(x)+\sum_{i=1}^{\epsilon}\left(u^{l_{\rho+i}}e_{\rho+i}(x)+
u^{4-l_{\rho+i}}e_{\rho+i+\epsilon}(x)\right)\right\rangle.$$

\vskip 2mm\par
  \textit{Therefore, the number of self-dual $(1+\alpha u^2)$-constacyclic codes over $R$ of length $n$ is equal to $5^\epsilon$}.

\vskip 3mm \noindent
  {\bf Proof.} Let $\mathcal{C}$ be any $(1+\alpha u^2)$-constacyclic code over $R$ of length $n$.
By Theorem 2.9 and its proof, we have that $\mathcal{C}=\bigoplus_{j=1}^r\langle u^{l_j}e_{j}(x)\rangle$
where $0\leq l_j\leq 4$ for all $j=1,\ldots,r$.
  By Corollary 4.1 and the proof of Theorem 2.9, we see that $\mathcal{C}^{\bot_E}=\bigoplus_{j=1}^r\langle u^{4-l_j}e_{\tau(j)}(x)\rangle$.
Since $\tau$ is a bijection on the set $\{1,\ldots,r\}$, we have $\mathcal{C}=\bigoplus_{j=1}^r\langle u^{l_{\tau(j)}}e_{\tau(j)}(x)\rangle$.
From this and by
Corollary 2.8(i), we deduce that $\mathcal{C}$ is self-dual if and only if $l_{\tau(j)}=4-l_j$ for all $j=1,\ldots,r$.
Then we have one of the following two cases.

\par
   (i) Let $1\leq j\leq \rho$. Then $\tau(j)=j$,
and $l_{\tau(j)}=4-l_j$ if and only
if $l_j=4-l_j$, which is equivalent to that $l_j=2$.

\par
   (ii) Let $j=\rho+i$, where $1\leq i\leq \epsilon$. In this case, $\tau(j)=j+\epsilon$ and $\tau(j+\epsilon)=j$.
Then $l_{\tau(j)}=4-l_j$ and $l_{\tau(j+\epsilon)}=4-l_{j+\epsilon}$, i.e., $l_{j+\epsilon}=4-l_j$ and $l_{j}=4-l_{j+\epsilon}$,
if and only if $l_{j+\epsilon}=4-l_j$.
\hfill $\Box$


\section{An example} \label{}
\noindent
  Let $\mathbb{F}_2=\{0,1\}$ and $R=\mathbb{F}_2[u]/\langle u^4\rangle=\mathbb{F}_2+u\mathbb{F}_2+u^2\mathbb{F}_2+u^3\mathbb{F}_2$
($u^4=0$). Then $R$ is a finite chain ring of $2^4=16$ elements.
   It is known that $x^7-1=x^7+1=f_1(x)f_2(x)f_3(x)$ where
\begin{center}
   $f_1(x)=x+1$, $f_2(x)=x^3+x+1$ and $f_3(x)=x^3+x^2+1$
\end{center}
are
irreducible polynomials in $\mathbb{F}_2[x]$. Obviously, $d_j={\rm deg}(f_j(x))$ where $d_1=1$ and $d_2=d_3=3$.

\par
   For $1\leq j\leq 3$, let $F_j(x)=\frac{x^7+1}{f_j(x)}$ and
find $g_j(x),h_j(x)\in \mathbb{F}_2[x]$ such that
$g_j(x)F_j(x)^2+h_j(x)f_j(x)^2=1.$
Then we calculate
$\varepsilon_j(x)=g_j(x)F_j(x)^2 \ ({\rm mod} \ x^{14}+1).$
Dividing $\varepsilon_j(x)$ by $x^7+1$, we obtain a unique pair $(e_{j,0}(x),e_{j,1}(x))$ of polynomials in
$\mathbb{F}_2[x]$ such that $\varepsilon_j(x)=e_{j,0}(x)+(x^7+1)e_{j,1}(x)$ and
${\rm deg}(e_{j,i}(x))\leq 6$ for $i=0,1$. Then we have
$$e_j(x)=e_{j,0}(x)+u^2e_{j,1}(x)\in R[x]/\langle x^7-(1+u^2)\rangle$$
by Corollary 2.8(ii). Precisely, we have

\par
  $e_1(x)={x}^{6}+ ( {u}^{2}+1 ) {x}^{5}+{x}^{4}+ ( {u}^{2}+1) {x}^{3}+{x}^{2}+ ( {u}^{2}+1 ) x+1$;

\par
  $e_2(x)={x}^{4}+{x}^{2}+ ( {u}^{2}+1 ) x+1$;

\par
  $e_3(x)={x}^{6}+ ( {u}^{2}+1 ) {x}^{5}+ ( {u}^{2}+1 ) {x}^{3}+1$,

\noindent
where $\tau(e_1(x))=e_1(x^{-1})=e_1(x)$ and $\tau(e_2(x))=e_2(x^{-1})=e_3(x)$ (mod $x^7-(1+u^2)$).
Hence $\rho=\epsilon=1$.

\vskip 3mm\par
   $\bullet$ By Theorem 2.9, there are $5^3=125$ distinct $(1+u^2)$-constacyclic codes over $R$ of length $7$:
$$\mathcal{C}_{(l_1,l_2,l_3)}=\left\langle g_{(l_1,l_2,l_3)}(x)
\right\rangle \ ({\rm mod} \ x^7-(1+u^2)),$$
where
$g_{(l_1,l_2,l_3)}(x)=u^{l_1}e_1(x)+u^{l_2}e_2(x)+u^{l_3}e_3(x)$, $0\leq l_1,l_2,l_3\leq 4.$
Moreover, the number of codewords contained in $\mathcal{C}_{(l_1,l_2,l_3)}$ is
equal to
$$|\mathcal{C}_{(l_1,l_2,l_3)}|
=2^{(4-l_1)+3(4-l_2)+3(4-l_3)}=2^{28-(l_1+3(l_2+l_3))}.$$

\par
   $\bullet$ By Theorems 4.2, there are $5$ distinct self-dual $(1+u^2)$-constacyclic codes over $R$ of
length $7$:
$$\mathcal{C}_{(2,l,4-l)}=\left\langle g_{(2,l,4-l)}(x)\right\rangle \ ({\rm mod} \ x^7-(1+u^2)), \ 0\leq l\leq 4,$$
where

\par
  $g_{(2,0,4)}(x)={u}^{2}{x}^{6}+{u}^{2}{x}^{5}+ ( {u}^{2}+1 ) {x}^{4}+{u}^{2
}{x}^{3}+ ( {u}^{2}+1 ) {x}^{2}+x+{u}^{2}+1$;

\par
  $g_{(2,1,3)}(x)= ( {u}^{3}+{u}^{2} ) {x}^{6}+ ( {u}^{3}+{u}^{2}
) {x}^{5}+ ( {u}^{2}+u ) {x}^{4}+ ( {u}^{3}+{u}
^{2} ) {x}^{3}+ ( {u}^{2}+u ) {x}^{2}+ ( {u}^{3}
+{u}^{2}+u ) x+{u}^{3}+{u}^{2}+u$;

\par
  $g_{(2,2,2)}(x)=u^2$;

\par
  $g_{(2,3,1)}(x)= ( {u}^{2}+u ) {x}^{6}+ ( {u}^{3}+{u}^{2}+u ) {x
}^{5}+ ( {u}^{3}+{u}^{2} ) {x}^{4}+ ( {u}^{3}+{u}^{2} ) {x}^{2}+ ( {u}^{3}+{u}^{2}+
u ) {x}^{3}+ ( {u}^
{3}+{u}^{2} ) x+{u}^{3}+{u}^{2}+u$;

\par
  $g_{(2,4,0)}(x)= ( {u}^{2}+1 ) {x}^{6}+{x}^{5}+{x}^{4}{u}^{2}+{x}^{3}+{x}^{
2}{u}^{2}+x{u}^{2}+{u}^{2}+1$.



\vskip 3mm \noindent {\bf Acknowledgments}
Part of this work was done when Yonglin Cao was visiting Chern Institute of Mathematics, Nankai University, Tianjin, China. Yonglin Cao would like to thank the institution for the kind hospitality. This research is
supported in part by the National Natural Science Foundation of
China (Grant Nos. 11471255).


\begin{thebibliography}{11}
\bibitem{s1} T. Abualrub, I. Siap, Cyclic codes over the ring $\mathbb{Z}_2+u\mathbb{Z}_2$ and $\mathbb{Z}_2+u\mathbb{Z}_2+u^2\mathbb{Z}_2$,
Des. Codes Cryptogr. {\bf 42} (2007), 273--287.

\bibitem{s2} M. Al-Ashker, M. Hamoudeh, Cyclic codes over $Z_2+uZ_2+u^2Z_2+\ldots+u^{k-1}Z_2$,
Turk. J. Math. {\bf 35}(4) (2011), 737--749.

\bibitem{s3} Y. Cao, On constacyclic codes over finite chain rings,
Finite Fields Appl. {\bf 24} (2013), 124--135.

\bibitem{s4} Y. Cao, Y. Gao, Repeate root cyclic $\mathbb{F}_q$-linear codes over
$\mathbb{F}_{q^l}$,
Finite Fields Appl. {\bf 31} (2015), 202--227.

\bibitem{s5} H. Q. Dinh, Constacyclic codes of length $p^s$ over
$\mathbb{F}_{p^m}+u \mathbb{F}_{p^m}$, J. Algebra, {\bf 324} (2010),
940--950.

\bibitem{s6}  S. T. Dougherty,  J-L. Kim,  H. Kulosman,  H. Liu: Self-dual
codes over commutative Frobenius rings, Finite Fields Appl. {\bf 16}, 14--26 (2010).

\bibitem{s7} M. Han, Y. Ye, S. Zhu, C. Xu, B. Dou,
Cyclic codes over $R = F_p + uF_p +\ldots+ u^{k-1}F_p$ with length $p^sn$,
Information Sciences, {\bf 181} (2011), 926--934.

\bibitem{s8} X. Kai, S. Zhu, P. Li, $(1+\lambda u)$-constacyclic codes over $\mathbb{F}_p[u]/\langle u^m\rangle$,
J. Franklin Inst. {\bf 347} (2010), 751--762.

\bibitem{s9} A. K. Singh, P. K. Kewat, On cyclic codes over the ring $\mathbb{Z}_p[u]/\langle u^k\rangle$,
Des. Codes Cryptogr. {\bf 72} (2015), 1--13.

\bibitem{s10} R. Sobhani, M. Esmaeili, Some constacyclic and codes over $\mathbb{F}_q[u]/\langle u^{t+1}\rangle$,
IEICE Trans. Fundam. Electron. {\bf 93} (2010), 808--813.

\bibitem{s11} R. Sobhani, Complete classification of $(\delta+\alpha u^2)$-constacyclic
codes of length $p^k$ over $\mathbb{F}_{p^m}+u\mathbb{F}_{p^m}+u^2\mathbb{F}_{p^m}$,
Finite Fields Appl. {\bf 34} (2015), 123--138.
\end{thebibliography}

\end{document}